\documentclass[pra,10pt,twocolumn,notitlepage,twoside,final,a4paper,
               showpacs]{revtex4}

\usepackage{amssymb}
\usepackage{amsmath}
\usepackage{ifthen}
\usepackage{graphicx}
\graphicspath{{./}{lines/}{pics/}}

\newcommand{\eq}[1]{\begin{math}#1\end{math}}

\newcommand{\eqxn}[1]{\begin{equation}#1\end{equation}}

\newcommand{\eqxna}[1]{\begin{eqnarray}#1\end{eqnarray}}

\newcommand{\ket}[1]{\left|#1\right\rangle}

\newcommand{\cc}[1]{#1^\star}
\newcommand{\abs}[1]{\left|#1\right|}
\newcommand{\sq}[1]{#1^2}
\newcommand{\asq}[1]{\sq{\abs{#1}}}
\newcommand{\mean}[1]{\left<#1\right>}
\newcommand{\dd}[1]{\frac{\mbox{d}}{\mbox{d}#1}}
\newcommand{\VC}{\tilde{C}}
\newcommand{\D}{\displaystyle}
\newcommand{\n}{\vspace{\jot}\\}

\newcommand{\lia}{\includegraphics*[bb=0 0 28 28, width=0.75cm, 
                  viewport=8 13 20 15]{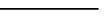}}
\newcommand{\lib}{\includegraphics*[bb=0 0 28 28, width=0.75cm, 
                  viewport=8 13 20 15]{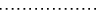}}
\newcommand{\lic}{\includegraphics*[bb=0 0 28 28, width=0.75cm, 
                  viewport=7 13 19 15]{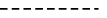}}
\newcommand{\lie}{\includegraphics*[bb=0 0 28 28, width=0.75cm, 
                  viewport=9.25 13 21 15]{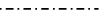}}
\newcommand{\lig}{\includegraphics*[bb=0 0 28 28, width=0.75cm, 
                  viewport=8.5 13 21 15]{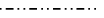}}
\newcommand{\lih}{\includegraphics*[bb=0 0 28 28, width=0.75cm, 
                  viewport=11 13 20 15]{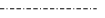}}
\newcommand{\sym}[1]{\includegraphics*[bb=0 0 28 28, width=0.75cm, 
                     viewport=8.25 13 20 16]{s#1.ps}}
\newcommand{\ls}[2]{\ifthenelse{\equal{#1}{0}}{\hspace{+0.75cm}}{}\nolinebreak%
                    \ifthenelse{\equal{#1}{1}}{\lia}{}\nolinebreak%
                    \ifthenelse{\equal{#1}{2}}{\lib}{}\nolinebreak%
                    \ifthenelse{\equal{#1}{3}}{\lic}{}\nolinebreak%
                    \ifthenelse{\equal{#1}{5}}{\lie}{}\nolinebreak%
                    \ifthenelse{\equal{#1}{7}}{\lig}{}\nolinebreak%
                    \ifthenelse{\equal{#1}{8}}{\lih}{}\nolinebreak%
                    \ifthenelse{\equal{#2}{0}}{}{\hspace{-0.75cm}\sym{#2}}}

\begin{document}

  \title{Entanglement and correlation in anisotropic
         quantum spin systems}
  \date{\today}
  \author{Ulrich Glaser}
  \affiliation{Integrated Systems Laboratory, 
               ETH Z\"urich, CH-8092 Z\"urich, Switzerland\\
               CL DAT LIB IO, Infineon Technologies, D-81541
               M\"unchen, Germany}
  \author{Helmut B\"uttner}
  \affiliation{Theoretische Physik I, 
               Universit\"at Bayreuth, D-95440 Bayreuth, Germany}
  \author{Holger Fehske}
  \affiliation{Institut f\"ur Physik, Theoretische Physik II, 
                Universit\"at Greifswald, D-17487 Greifswald, Germany}
  \begin{abstract}
    Analytical expressions for the entanglement measures concurrence, 
    i-concurrence and 3-tangle in terms of spin correlation functions 
    are derived using general symmetries of the quantum spin system. 
    These relations  are exploited for the one-dimensional 
    XXZ-model, in particular the concurrence and the critical temperature 
    for disentanglement are calculated for finite systems with up 
    to six qubits. A recent NMR quantum error correction experiment is 
    analyzed within the framework of the proposed theoretical approach.
  \end{abstract}


  \maketitle


\section{Introduction}
\label{k_Introduction_Basics}

  Quantum entanglement was already pointed out by Schr\"odinger 
  \cite{Schroedinger:1935x} to be a crucial element of quantum mechanics. 
  Research was refocused on quantum entanglement in the last fifteen years
  because the field of quantum information theory 
  (cf.~\cite{Macchiavello:2000, Nielsen:2001}) developed rather
  quickly. 
  Recent papers concerning entanglement in quantum spin systems address 
  questions about the maximum entanglement of nearest neighbor qubits 
  belonging to a ring of \eq{N} qubits in a translationally invariant 
  quantum state \cite{Connor:2001}, the dependence of entanglement between 
  two spins on temperature, external magnetic field strength and/or
  anisotropy for the one-dimensional isotropic Heisenberg model 
  \cite{Arnesen:2001, Wang:2001_12, Wang:2002_3, Wang:2002_8, Wang:2002_10, 
  Schliemann:2002}, Ising model \cite{Gunlycke:2001}, 
  the three-qubits XXZ-model \cite{Wang:2001_12} 
  and the XY-model \cite{Kamta:2002}. 
  Further topics are entanglement close to quantum phase 
  transitions \cite{Wang:2001_12, Osterloh:2002, Osborne:2002, Bose:2002,
  Vidal:2002, Latorre:2003} and 
  global entanglement with an application to quantum error correction code 
  subspaces \cite{Meyer:2002}.

  In the present paper, several new aspects of quantum entanglement 
  are discussed, in particular how the various measures of entanglement 
  can be related to correlation functions. After introducing briefly
  the basic notations and definitions in the next Section,  
  the functional dependences of the entanglement measures 
  concurrence~\cite{Hill:1997,Wootters:1998}, 
  i-concurrence \cite{Rungta:2001} (in small systems) and 
  3-tangle \cite{Coffman:2000} on spin correlation functions 
  (including spin expectation values) are established in 
  Sec.~\ref{k_Entanglement_ExpectationValues}. 
  Necessary and sufficient conditions for a positive concurrence are found.  
  In Sec.~\ref{k_XXZ-Model}, the expectation values, correlation functions 
  and concurrence of both ground and excited states of the 
  one-dimensional XXZ-model as well as the mixed state of the quantum
  system at finite temperature are calculated 
  analytically in terms of the eigenenergies.
  The concurrence of a \eq{N = 4} quantum spin system  and 
  the critical temperature where the concurrence vanishes 
  are examined in detail. Results are also presented  
  for \eq{N=2}, \eq{3}, \eq{5} and \eq{6} qubit systems. 
  Finally, the entanglement of a quantum system with 
  \eq{N = 5} qubits in a NMR quantum error correction 
  experiment~\cite{Knill:2001} is discussed and partly quantified 
  in terms of the entanglement measures in
  Sec.~\ref{k_Experiment}.

\section{Basic notations}
\label{notations}

  Consider a quantum system consisting of \eq{N} qubits on numbered sites. 
  The basis of the state of one qubit is given by 
  \eq{\ket{0}}, \eq{\ket{1}} which are the eigenstates of \eq{\sigma^z} 
  (\eq{\sigma^{x}}, \eq{\sigma^{y}}, \eq{\sigma^{z}} 
  denote the Pauli spin operators) with eigenvalues \eq{-1}, \eq{ +1}, 
  respectively. An unentangled state of \eq{N} 
  qubits is the direct product of the single qubits, e.g., 
  \eq{\ket{\psi}_{12 \cdots N} = \ket{0}_{1} \otimes \ket{0}_{2} 
  \otimes \cdots \otimes \ket{0}_{N} =: \ket{00 \cdots 0}_{12 \cdots N}}. 
  If unambiguous then indices indicating site numbers will be omitted
  in the following because the qubits are arranged with increasing site 
  number. Thus site information is contained in the ordering of the qubits. 
  The Hamiltonian \eq{H}  and the density operator \eq{\rho} 
  describing such quantum spin systems are usually expressed 
  in terms of the identity operator \eq{I}, the Pauli spin operators, 
  and/or the operators \eq{\sigma^{\pm} := \frac{1}{2} 
  \left( \sigma^{x} \pm i \sigma^{y} \right)}.

  The state of the spin system becomes mixed at finite temperatures. 
  The operator representing this state is frequently 
  called thermal density operator. In thermodynamical equilibrium, 
  it is given by the operator  
  \eq{\label{g_rhoT} \rho = Z^{-1} \exp [-\beta H] },
  where \eq{\beta = (k_B T)^{-1}}, \eq{k_B} denotes the Boltzmann constant, 
  \eq{T} is the temperature of the system and 
  \eq{Z = \text{Tr}\exp [-\beta H] } is the partition function.

  Spin expectation values and correlation functions are defined as 
  \begin{equation}
    K_{n \cdots m}^{\nu \cdots \mu} := 
      \mean{\sigma_{n}^{\nu} \cdots \sigma_{m}^{\mu}} = 
      \text{Tr} \left( \rho \,\sigma_{n}^{\nu} \cdots \sigma_{m}^{\mu} 
    \right)\,,
  \end{equation}
  where \eq{n, \ldots, m \in \left\{ 1, \ldots, N \right\} } 
  and \eq{\nu, \ldots, \mu \in \left\{ x, y, z, +, - \right\} } specify 
  qubit and operator, respectively. 
  Furthermore, in what follows, the  
  \eq{z}-component of the total spin operator
  \eq{\label{g_Sz}
       S^z := \sum_{n=1}^{N} \sigma_{n}^{z},} 
  the spinflip operator 
  \eq{\label{g_F}
       F := \bigotimes_{n=1}^{N} \sigma_{n}^{x}}
  and, assuming periodic boundary conditions, the translation operator 
  \eq{T(l) := \exp[-ilaK]} with the lattice constant \eq{a} and the 
  momentum operator \eq{\hbar K = P} will be used occasionally.

\section{Entanglement and Correlation Functions}
\label{k_Entanglement_ExpectationValues}

  The functional dependence of entanglement (measured in terms of the 
  concurrence, i-concurrence and 3-tangle) on correlation functions  of 
  the operators \eq{\sigma^{x}}, \eq{\sigma^{y}}, \eq{\sigma^{z}}, 
  \eq{\sigma^{\pm}} is now discussed as far as possible  
   without an explicit specification of the model Hamiltonian.

  Using the basis \eq{\ket{0}} and \eq{\ket{1}},
  the expansion coefficients of the (reduced) density operator of one qubit 
  \eq{n} (\eq{1 \le n \le N})  are given by spin expectation values only:
  \begin{subequations}
    \label{g_rho1_sym}
    \eqxna{\rho_{11}^{(1)} 
             &=& \frac{1}{2} \left( 1 - K_{n}^{z} \right), \\
           \rho_{22}^{(1)} 
             &=& \frac{1}{2} \left( 1 + K_{n}^{z} \right), \\
           \rho_{12}^{(1)} = \cc{\left( \rho_{21}^{(1)} \right)}
             &=& K_{n}^{+} = \cc{\left( K_{n}^{-} \right)}.}
  \end{subequations}

  In the same manner, the (reduced) density operator of 
  two qubits \eq{n} and \eq{m} (\eq{1 \le n < m \le N}) can be expressed  
  in the basis \eq{\ket{00}}, \eq{\ket{01}}, \eq{\ket{10}} and 
  \eq{\ket{11}}. If the Hamiltonian  
  commutes with the \eq{z}-component of the total spin operator, 
  the corresponding expressions can be simplified, yielding 
  \begin{subequations}
    \label{g_rho2_sym}
    \eqxna{\rho_{11}^{(2)} 
             &=& \frac{1}{4} \left( 1 - K_{n}^{z} - K_{m}^{z} +  
                                    K_{nm}^{zz} \right), \\
           \rho_{22}^{(2)} 
             &=& \frac{1}{4} \left( 1 - K_{n}^{z} + K_{m}^{z} -  
                                    K_{nm}^{zz} \right), \\
           \rho_{33}^{(2)} 
             &=& \frac{1}{4} \left( 1 + K_{n}^{z} - K_{m}^{z} -  
                                    K_{nm}^{zz} \right), \\
           \rho_{44}^{(2)} 
             &=& \frac{1}{4} \left( 1 + K_{n}^{z} + K_{m}^{z} +  
                                    K_{nm}^{zz} \right), \\
           \rho_{23}^{(2)} = \cc{\left( \rho_{32}^{(2)} \right)}
             &=& K_{nm}^{+-} = \cc{\left( K_{nm}^{-+} \right)},}
  \end{subequations}
  and all other coefficients are equal zero.

  {\em Concurrence \eq{C}} has been introduced by Wootters 
  \cite{Wootters:1998} as a measure to quantify entanglement. 
  Let \eq{\rho} be the density operator representing a pure or mixed state of
  two qubits \eq{n} and \eq{m}. Then 
  \eqxna{\label{g_C}
         C_{nm} 
           &=& \max \left( 0, \VC_{nm} \right), \\
         \VC_{nm} 
           &=& 2 \lambda_{max} - \sum\limits_{j=1}^{4} \lambda_{j},}
  where \eq{\lambda_{max} := \max \left( \lambda_{1}, \lambda_{2}, 
            \lambda_{3}, \lambda_{4} \right) } and 
  \eq{\lambda_{1}}, \eq{\lambda_{2}}, \eq{\lambda_{3}}, \eq{\lambda_{4}} 
  are the non-negative, real eigenvalues of the matrix 
  \eq{R = \sqrt{\rho \left( \sigma^{y} \otimes \sigma^{y} \right) 
                \cc{\rho} \left( \sigma^{y} \otimes \sigma^{y} \right) }.}

  For a density operator with the coefficients (\ref{g_rho2_sym}), 
  one has
  \begin{subequations} 
    \label{g_lambda} 
    \eqxna{\lambda_{1} = \lambda_{2} 
             &=& \frac{1}{4} \xi^{+}, \\
           \lambda_{3,4} 
             &=& \frac{1}{4} \abs{\xi^{-} \pm 4 \abs{K_{nm}^{+-}}}, \\
           \xi^{\pm}
             &=& \sqrt{\left( 1 \pm K_{nm}^{zz} \right)^2 -
                       \left( K_{n}^{z} \pm K_{m}^{z} \right)^2}, }
  \end{subequations}
  \eqxna{\label{g_CK}
        \VC_{nm} = \left\{
        \begin{array}{ll}
          \D \frac{1}{2} \left( 4 \abs{K_{nm}^{+-}} - \xi^{+} \right)
            & \D \text{ if } 
                 \lambda_{1} = \lambda_{2} < \lambda_{3} \\
            & \D \text{ and } \xi^{-} > 4 \abs{K_{nm}^{+-}}, \n
          \D \frac{1}{2} \left( \xi^{-} - \xi^{+} \right)
            & \D \text{ if } 
                 \lambda_{1} = \lambda_{2} < \lambda_{3} \\
            & \D \text{ and } \xi^{-} \le 4 \abs{K_{nm}^{+-}}, \n
          \D - \frac{1}{2} \xi^{-}
            & \D \text{ if } 
                 \lambda_{1} = \lambda_{2} \ge \lambda_{3}, \lambda_{4} \\
            & \D \text{ and } \xi^{-} > 4 \abs{K_{nm}^{+-}}, \n
          \D -2 \abs{K_{nm}^{+-}}
            & \D \text{ if } 
                 \lambda_{1} = \lambda_{2} \ge \lambda_{3}, \lambda_{4} \\
            & \D \text{ and } \xi^{-} \le 4 \abs{K_{nm}^{+-}}. \n
        \end{array} \right.}
  Thus Eqs.~(\ref{g_C}) and (\ref{g_CK}) yield the functional 
  dependence of the concurrence on correlation functions using 
  \eq{S^{z}}-symmetry only.
  
  Cases \eq{3} and \eq{4} of Eq.~(\ref{g_CK}) are not interesting 
  because \eq{\VC_{nm} \le 0} and thus \eq{C_{nm} = 0}. 
  With the help of cases 1 and 2, it is straightforward 
  to find the following necessary and sufficient conditions 
  for entanglement,
  \eqxna{\label{g_nec} & K_{nm}^{zz} - K_{n}^{z} K_{m}^{z} < 0, \\
         \label{g_suf} & K_{nm}^{zz} - K_{n}^{z} K_{m}^{z} < 0 \text{ and }
                         \xi^{+} < 4 \abs{K_{nm}^{+-}},}
  respectively. These results are similar to the conjecture 
  that the ground 
  state of the transverse Ising-model and the XY-model is entangled, 
  iff, according to \cite{Osborne:2002}, 
  \eq{K_{nm}^{\mu \nu} - K_{n}^{\mu} K_{m}^{\nu} \not= 0}.

  Eqs.~(\ref{g_nec}) and (\ref{g_suf}) can be interpreted in the following
  way: If the state of two qubits in a system with
  \eq{K_{n}^{z} = 0} and/or \eq{K_{m}^{z} = 0} is entangled then the 
  z-components of the spins must be correlated antiferromagnetically.
  The maximal entangled states are the two Bell-states
  \eq{\ket{\psi^\pm} = \frac{1}{\sqrt{2}} 
                       \left( \ket{01} \pm \ket{10} \right) }. 
  If \eq{K_{n}^{z} K_{m}^{z} > 0}, e.g., if an appropriate external magnetic
  field is applied, entanglement of qubits with ferromagnetically correlated
  z-components of the spins is possible. 
  The sufficient condition requires moreover that the correlations of the 
  two qubits need to be greater than a minimum value to create entanglement. 
  Again an appropriate external magnetic field reduces this demand.

  If the system exhibits additional spin flip symmetry, 
  \eq{K_{n}^{z} = K_{m}^{z} = 0} and \eq{K_{nm}^{+-} = K_{nm}^{-+}} 
  result.
  Then Eqs.~(\ref{g_rho2_sym}), (\ref{g_lambda}) 
  and (\ref{g_CK}) simplify and case \eq{1} of Eq.~(\ref{g_CK}) 
  coincides to the result published in \cite{Wang:2002_8}. 
  Necessary and sufficient conditions for entanglement are now
  \eqxna{& K_{nm}^{zz} < 0, \\
         & K_{nm}^{zz} < 0 \text{ and }
           1 < \abs{K_{nm}^{xx}} + \abs{K_{nm}^{yy}} + \abs{K_{nm}^{zz}},}
  respectively. Here the relation 
  \eq{K_{nm}^{xx} = K_{nm}^{yy} = 2 K_{nm}^{+-}}, which is correct because 
  of \eq{S^z} and \eq{F} symmetry, was used. 

  {\em I-concurrence \eq{\bar{C}}} has been proposed by 
  Rungta {\it et al.}~\cite{Rungta:2001} as an
  entanglement measure. Let 
  \eq{AB} be a quantum system consisting of two subsystems 
  \eq{A} and \eq{B} 
  with dimensions \eq{d_A} and \eq{d_B}, respectively. The density 
  operators representing the state of these systems are denoted 
  \eq{\rho_{AB}}, \eq{\rho_A} and
  \eq{\rho_B}, respectively. If \eq{\rho_{AB}} represents a pure state then 
  the entanglement of this state with respect to the two subsystems \eq{A} 
  and \eq{B} is quantified by 
  \eqxn{\label{g_IC}
        \bar{C}_{A-B} = \sqrt{2 \left[ 1 - \text{Tr} 
                                       \left( \rho_A^2 \right) \right] },}
  where \eq{\rho_A = \text{Tr}_B \left( \rho_{AB} \right) } is the reduced
  density operator of subsystem \eq{A}. It is known from 
  \cite{Rungta:2001} that \eq{0 \le \bar{C}_{A-B} \le \sqrt{2 \frac{d-1}{d}}}, 
  where \eq{d = \min \left( d_A, d_B \right) }. A different notation is 
  occasionally used for qubits: For example \eq{\bar{C}_{12-34}} denotes the 
  entanglement of the state where subsystems \eq{A} and \eq{B} consist of 
  qubits \eq{1}, \eq{2} and \eq{3}, \eq{4}, respectively. Note that
  \eq{C_{nm} = \bar{C}_{n-m}} if the state of qubits \eq{n} and \eq{m} is pure.

  From Eqs.~(\ref{g_rho1_sym}) and (\ref{g_IC}), it follows that 
  \eqxn{\label{g_IC_1r}
        \bar{C}_{n-rest} = \sqrt{1 - (K_n^z)^2 - 4 K_n^+ K_n^-}.} 
  If the Hamiltonian commutes with \eq{S^z}, 
  Eqs.~(\ref{g_rho2_sym}) and (\ref{g_IC}) yield
  \eqxna{\label{g_IC_2r}
           & & \hspace{-0.5cm} \bar{C}_{nm-rest} = \\
           & & \sqrt{\frac{3}{2} - 
               \frac{1}{2} \left[ (K_{nm}^{zz})^2 +
                                  (K_{n}^{z})^2 + 
                                  (K_{m}^{z})^2 \right] -
               4 \asq{K_{nm}^{+-}}}. \nonumber}
  In an analogous way the i-concurrence of three and more qubits 
  can be expressed in terms of correlation functions.

  Two highly entangled qubits cannot be much entangled with 
  the remaining system and vice versa. This property is ensured in 
  Eqs.~(\ref{g_IC_1r}) and (\ref{g_IC_2r}). They indicate high entanglement 
  in the system if the absolute values of expectation values and 
  correlation functions are as small as possible (preferable zero). This is
  contrary to the requirements for a high concurrence. 

  {\em 3-tangle \eq{\tau}} has been suggested by 
  Coffman {\it et al.}~\cite{Coffman:2000} to quantify the entanglement 
  of a pure state of three qubits \eq{1}, \eq{2} 
  and \eq{3} in the following way:
  \eqxn{\label{g_3tangle}
        \tau_{123} = C_{1-23}^2 - C_{12}^2 - C_{13}^2,}
  where \eq{C_{1-23}^2 = 4 \det (\rho_1) = \bar{C}_{1-23}^2} and
  \eq{\rho_1 = \text{Tr}_{23} (\rho_{123})}. Note that
  \eq{\tau_{123}} does not contain the entanglement of two out of the three 
  qubits and \eq{\tau_{123}} does not depend on the arbitrary choice of qubit
  \eq{1} as the ''central`` qubit.
 
  The 3-tangle \eq{\tau_{123}} can be expressed in terms of correlation   
  functions if the Hamiltonian of the system commutes with \eq{S^z}. 
  This is achieved by 
  expressing the right hand side of Eq.~(\ref{g_3tangle}) in terms of 
  correlation functions with the help of Eqs.~(\ref{g_C}), (\ref{g_CK}) 
  and (\ref{g_IC_1r}).

\section{XXZ-Model}
\label{k_XXZ-Model}

  The Hamiltonian \eq{H(J, \Delta)} of the one-dimensional 
  (spatial) homogeneous
  XXZ-model reads (cf.~\cite{Takahashi:1999}) 
  \eqxn{\label{g_H}
        \begin{array}{rcl}
          H
          &=& \D \frac{1}{2} J \sum\limits_{n=1}^{N} 
                 \left( \sigma_{n}^{+} \sigma_{n+1}^{-} + 
                        \sigma_{n}^{-} \sigma_{n+1}^{+} + 
                        \frac{1}{2} \Delta \sigma_{n}^{z} \sigma_{n+1}^{z} 
                 \right).
        \end{array}}
  The coupling constant \eq{J} specifies the 
  strength of nearest-neighbor spin interaction. 
  Anisotropy in spin space is quantified by \eq{\Delta}. 
  Periodic boundary conditions are assumed. 
  In what follows, all energies are measured in units of $J$.

  The XXZ-model possesses some interesting symmetries. 
  The Hamiltonian (\ref{g_H}) commutes with the 
  \eq{z}-component of the total spin operator \eq{S^z}, 
  the spinflip operator \eq{F} and the translation operator \eq{T(l)}.
  Unfortunately \eq{S^{z}} and \eq{F} do not commute but of
  course it is possible to classify eigenstates of \eq{H} 
  by eigenvalues \eq{s} of \eq{S^{z}} 
  and eigenvalues \eq{k} of \eq{\frac{i N}{2 \pi} \ln[T(1)]}.
  Because of \eq{F}-symmetry, it is sufficient to solve the eigenvalue
  problem of \eq{H} in subspace with \eq{s \le 0}. 
 
  It was shown in \cite{Yang:1966} that \eq{H(J, \Delta)} and 
  \eq{H(-J, -\Delta)} possess for even \eq{N} a spectrum of 
  identical eigenvalues in each subspace of \eq{s} because the operator 
  \eq{A := \bigotimes_{n=1,3,\ldots}^{N-1} \sigma_{n}^{z}} 
  commutes with \eq{S^z} and \eq{A H(J, \Delta) A^{-1} = 
  H(-J, -\Delta) = -H(J, -\Delta)}. 

  Some correlation functions of the XXZ model are interdependent.
  If only eigenstates with
  equal \eq{s} participate in the thermal density operator then it is
  straightforward to show that
  \eqxn{K_{m_{1}^{1} \cdots m_{1}^{\xi_1}}^{z \cdots z} =
          (-1)^{(\frac{N}{2} - s)} 
          K_{m_{2}^{1} \cdots m_{2}^{\xi_{2}}}^{z \cdots z}, }
  where \eq{m_{1}^{1}, \ldots, m_{1}^{\xi_1}} and 
  \eq{m_{2}^{1}, \ldots, m_{2}^{\xi_2}} are the elements of \eq{M_1} 
  and \eq{M_2}, respectively, \eq{\xi_1 + \xi_2 = N}, 
  \eq{M_1 \cup M_2 = \{ 1, 2, \ldots, N \} } and 
  \eq{M_1 \cap M_2 = \varnothing}. 

  If $H$ has \eq{S^z}- and \eq{F}-symmetry, only \eq{K^{zz}_{nm}} and 
  \eq{K^{+-}_{nm}} appear in Eq.~(\ref{g_CK}). These correlation functions
  can be expressed in terms of the partition function.
  For example, \eq{K^{zz}_{n(n+1)}} and 
  \eq{K^{+-}_{n(n+1)}} read (cf. \cite{Orbach:1958})
  \eqxna{\label{g_Ring_Kzz}
         K^{zz}_{n(n+1)} 
           & = & - \frac{4}{N J \beta} \dd{\Delta} \ln Z, \\
         \label{g_Ring_K+-}
         K^{+-}_{n(n+1)}
           & = & - \frac{1}{N \beta} 
                 \left( \dd{J} -
                        \frac{\Delta}{J} \dd{\Delta} \right) \ln Z.}
  Using these relations, the correlation functions and concurrence of the 
  eigenstates and the thermal state of nearest neighbor 
  qubits can be calculated by knowing only the eigenvalues of the Hamiltonian.
  It is straightforward to express further expectation values and correlation
  functions in terms of the
  partition function using the same method. Possibly, the Hamiltonian has 
  to be supplemented (e.g.~adding to \eq{H} appropriate
  external magnetic field terms yields \eq{K^{z}_{n}} again as derivatives 
  of \eq{\ln Z}). 

  As another application of eq.~(\ref{g_CK}), the concurrence of nearest 
  neighbor qubits of the ground state in the anisotropic XXZ-model 
  with \eq{J = -1,\, \Delta = -\frac{1}{2}} and an odd number of
  qubits is considered.
  It is known from \cite{Stroganov:2001} that
  \eq{K_{n(n+1)}^{zz} = -\frac{1}{2} + \frac{3}{2 N^2}} and
  \eq{K_{n(n+1)}^{+-} = \frac{1}{2} K_{n(n+1)}^{xx} = 
      \frac{1}{2} K_{n(n+1)}^{yy} = \frac{5}{16} + \frac{3}{16 N^2}}. 
  Therefore \eq{K_{n(n+1)}^{zz} < 0} and
  \eq{1 < \abs{K_{n(n+1)}^{xx}} + \abs{K_{n(n+1)}^{yy}} + 
          \abs{K_{n(n+1)}^{zz}}} 
  for \eq{N \ge 3}. Thus the concurrence is 
  \eq{C_{n(n+1)} = \frac{3}{8} \left( 1 - \frac{1}{N^2} \right) }. 
  Concurrence is increasing with odd \eq{N} whereas 
  the concurrence of nearest neighbor qubits of the ground
  state in the isotropic antiferromagnetic Heisenberg model 
  decreases with increasing even \eq{N} in 
  all cases that have been calculated by 
  O'Connor {\em et al.}~\cite{Connor:2001}.

  Now the XXZ model is considered on a finite chain.
  Of course, the calculation of eigenstates and eigenvalues is getting
  more involved with increasing \eq{N} in general. Therefore, in what 
  follows, only small spin chains with \eq{2 \le N \le 6} are considered.

  For the case $N=4$, the eigenstates \eq{\ket{\psi}} 
  are given in Table \ref{t_Ring_4} together with \eq{C_{n(n+1)}} 
  and \eq{C_{n(n+2)}}, i.e., the entanglement of 
  nearest and next-to-nearest neighbor qubits in these eigenstates
  measured in terms of concurrence (\ref{g_C}).%
  \begin{table*}[t]
    \sffamily
    \caption[Eigenstates of $H$ for $N = 4$]
            {\label{t_Ring_4}
             Classification of the eigenstates of the XXZ-model (\eq{N = 4}) 
             and concurrence of nearest and next-to-nearest neighbor qubits.
             Normalization factors are given as 
             \eq{\eta_{1,2} := \sqrt{4 + 2 ( \mu_{1,2} )^2 }}, where
             \eq{\mu_{1,2} := -\frac{1}{2} \Delta 
                 \mp \frac{1}{2} \sqrt{\Delta^{2} + 8}}.
            }
    \centering
    \newcommand{\pf}{\rule[-1.0ex]{0mm}{3.5ex}}
    \begin{tabular}[t]{|c|c|c|c|c|c|}
      \hline
      \pf \eq{s} 
        & \eq{k}
        & \eq{E}
        & \eq{\ket{\psi}}
        & \eq{C_{n(n+1)}}
        & \eq{C_{n(n+2)}} \\
      \hline
      \hline
      \pf \eq{-2}
        & \eq{0}
        & \eq{\Delta}
        & \eq{\ket{0000}}
        & \eq{0}
        & \eq{0} \\
      \hline
      \pf \eq{-1}
        & \eq{0}
        & \eq{1}
        & \eq{\frac{1}{2} \left( \ket{1000} + \ket{0100} + 
                                 \ket{0010} + \ket{0001} \right)}
        & \eq{\frac{1}{2}}
        & \eq{\frac{1}{2}} \\
      \pf \eq{-1}
        & \eq{1}
        & \eq{0}
        & \eq{\frac{1}{2} \left( \ket{1000} + i \ket{0100} - 
                                 \ket{0010} - i \ket{0001} \right)}
        & \eq{\frac{1}{2}}
        & \eq{\frac{1}{2}} \\
      \pf \eq{-1}
        & \eq{2}
        & \eq{-1}
        & \eq{\frac{1}{2} \left( \ket{1000} - \ket{0100} + 
                                 \ket{0010} - \ket{0001} \right)}
        & \eq{\frac{1}{2}}
        & \eq{\frac{1}{2}} \\
      \pf \eq{-1}
        & \eq{3}
        & \eq{0}
        & \eq{\frac{1}{2} \left( \ket{1000} - i \ket{0100} - 
                                 \ket{0010} + i \ket{0001} \right)}
        & \eq{\frac{1}{2}}
        & \eq{\frac{1}{2}} \\
      \hline
      \pf \eq{0}
        & \eq{0}
        & \eq{\mu_1}
        & \eq{\frac{1}{\eta_1} 
              \left( \ket{1100} + \ket{0110} + \ket{0011} + \ket{1001} +
                     \mu_1 \ket{1010} + \mu_1 \ket{0101} \right)}
        & \eq{\max \left\{ 0, \frac{- 2 \mu_1 - 1}{2 + (\mu_1)^2} \right\} }
        & \eq{\max \left\{ 0, \frac{2 - (\mu_1)^2}{2 + (\mu_1)^2} \right\} } \\
      \pf \eq{0}
        & \eq{0}
        & \eq{\mu_2}
        & \eq{\frac{1}{\eta_2} 
              \left( \ket{1100} + \ket{0110} + \ket{0011} + \ket{1001} +
                     \mu_2 \ket{1010} + \mu_2 \ket{0101} \right)}
        & \eq{\max \left\{ 0, \frac{2 \mu_2 - 1}{2 + (\mu_2)^2} \right\} }
        & \eq{\max \left\{ 0, \frac{2 - (\mu_2)^2}{2 + (\mu_2)^2} \right\} } \\
      \pf \eq{0}
        & \eq{1}
        & \eq{0}
        & \eq{\frac{1}{2} \left( \ket{1100} + i \ket{0110} - 
                                 \ket{0011} - i \ket{1001} \right)}
        & \eq{0}
        & \eq{1} \\
      \pf \eq{0}
        & \eq{2}
        & \eq{0}
        & \eq{\frac{1}{2} \left( \ket{1100} - \ket{0110} + 
                                 \ket{0011} - \ket{1001} \right)}
        & \eq{0}
        & \eq{1} \\
      \pf \eq{0}
        & \eq{2}
        & \eq{- \Delta}
        & \eq{\frac{1}{\sqrt{2}} \left( \ket{1010} - \ket{0101} \right)}
        & \eq{0}
        & \eq{0} \\
      \pf \eq{0}
        & \eq{3}
        & \eq{0}
        & \eq{\frac{1}{2} \left( \ket{1100} - i \ket{0110} - 
                                 \ket{0011} + i \ket{1001} \right)}
        & \eq{0}
        & \eq{1} \\
      \hline
    \end{tabular}
  \end{table*}
  Eigenstates with
  \eq{s > 0} are obtained by applying \eq{F} on eigenstates with \eq{s < 0}.%

  The partition function, correlation functions 
  and concurrences at finite temperatures are calculated as%
  \begin{widetext}
  \eqxna{Z 
           &=& 2 \zeta^{-\Delta} + \zeta^{\Delta} + 2 \zeta^{-1} + 
               2 \zeta + 7 + 
               \zeta^{-\mu_1} + \zeta^{-\mu_2}, \\
         K_{n(n+1)}^{zz}
           &=& \frac{1}{Z} \left( 
               2 \zeta^{-\Delta} - \zeta^{\Delta} -
               \frac{(\mu_1)^2}{2 + (\mu_1)^2} \zeta^{-\mu_1} -
               \frac{(\mu_2)^2}{2 + (\mu_2)^2} \zeta^{-\mu_2} \right), \\
         K_{n(n+1)}^{+-}
           &=& \frac{1}{Z} \left( 
               \frac{1}{2} \zeta^{-1} - \frac{1}{2} \zeta +
               \frac{\mu_1}{2 + (\mu_1)^2} \zeta^{-\mu_1} +
               \frac{\mu_2}{2 + (\mu_2)^2} \zeta^{-\mu_2} \right), \\
         C_{n(n+1)}
           &=& \max \left\{ 0, \frac{1}{Z} \left( 
               \abs{\zeta^{-1} - \zeta +
                    \frac{2 \mu_1}{2+(\mu_1)^2} \zeta^{-\mu_1} +
                    \frac{2 \mu_2}{2+(\mu_2)^2} \zeta^{-\mu_2}} 
                    \right. \right. - \nonumber \\
           & & \qquad \qquad \qquad \left. \left.
               \abs{2 \zeta^{-\Delta} + \zeta^{-1} + \zeta + \frac{7}{2} + 
                    \frac{1}{2+(\mu_1)^2} \zeta^{-\mu_1} +
                    \frac{1}{2+(\mu_2)^2} \zeta^{-\mu_2}}
               \right) \right\}, \makebox[1.5cm]{} \\
         K_{n(n+2)}^{zz}
           &=& \frac{1}{Z} \left(
               2 \zeta^{-\Delta} + \zeta^{\Delta} - 3 +
               \frac{(\mu_1)^2 - 2}{2 + (\mu_1)^2} \zeta^{-\mu_1} +
               \frac{(\mu_2)^2 - 2}{2 + (\mu_2)^2} \zeta^{-\mu_2} \right), \\
         K_{n(n+2)}^{+-}
           &=& \frac{1}{Z} \left(
               \frac{1}{2} \zeta^{-1} + \frac{1}{2} \zeta - \frac{3}{2} +
               \frac{1}{2 + (\mu_1)^2} \zeta^{-\mu_1} +
               \frac{1}{2 + (\mu_2)^2} \zeta^{-\mu_2} \right), \\
         C_{n(n+2)}
           &=& \max \left\{ 0, \frac{1}{Z} \left(
               \abs{\zeta^{-1} + \zeta - 3 +
                    \frac{2}{2+(\mu_1)^2} \zeta^{-\mu_1} +
                    \frac{2}{2+(\mu_2)^2} \zeta^{-\mu_2}} \right. \right. - 
               \nonumber \\
           & & \qquad \qquad \qquad \left. \left.
               \abs{2 \zeta^{-\Delta} + \zeta^{\Delta} + \zeta^{-1} + 
                    \zeta + 2 + 
                    \frac{(\mu_1)^2}{2+(\mu_1)^2} \zeta^{-\mu_1} +
                    \frac{(\mu_2)^2}{2+(\mu_2)^2} \zeta^{-\mu_2}}
               \right) \right\}, }
  \end{widetext}%
  where \eq{\zeta := e^{\beta J}} and 
  \eq{\mu_{1,2} := -\frac{1}{2} \Delta \mp \frac{1}{2} \sqrt{\Delta^{2} + 8}}. 

\begin{figure}[b]
    \sffamily
    \vspace*{-0.5cm}
    \begingroup%
      \makeatletter%
      \newcommand{\GNUPLOTspecial}{%
        \@sanitize\catcode`\%=14\relax\special}%
      \setlength{\unitlength}{0.1bp}%
    \begin{picture}(2700,2592)(0,0)%
    \includegraphics{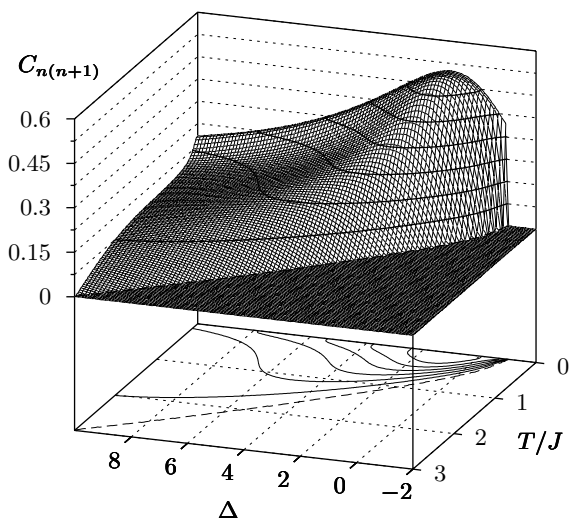}
    \put(631,518){\makebox(0,0){$8$}}%
    \put(843,494){\makebox(0,0){$6$}}%
    \put(1056,470){\makebox(0,0){$4$}}%
    \put(1268,445){\makebox(0,0){$2$}}%
    \put(1480,421){\makebox(0,0){$0$}}%
    \put(1693,397){\makebox(0,0){$-2$}}%
    \put(268,1998){\makebox(0,0)[l]{$C_{n(n+1)}$}}%
    \put(2151,581){\makebox(0,0)[l]{$T / J$}}%
    \put(1021,331){\makebox(0,0)[l]{$\Delta$}}%
    \put(631,518){\makebox(0,0){$8$}}%
    \put(843,494){\makebox(0,0){$6$}}%
    \put(1056,470){\makebox(0,0){$4$}}%
    \put(1268,445){\makebox(0,0){$2$}}%
    \put(1480,421){\makebox(0,0){$0$}}%
    \put(1693,397){\makebox(0,0){$-2$}}%
    \put(268,1998){\makebox(0,0)[l]{$C_{n(n+1)}$}}%
    \put(2151,581){\makebox(0,0)[l]{$T / J$}}%
    \put(1021,331){\makebox(0,0)[l]{$\Delta$}}%
    \put(631,518){\makebox(0,0){$8$}}%
    \put(843,494){\makebox(0,0){$6$}}%
    \put(1056,470){\makebox(0,0){$4$}}%
    \put(1268,445){\makebox(0,0){$2$}}%
    \put(1480,421){\makebox(0,0){$0$}}%
    \put(1693,397){\makebox(0,0){$-2$}}%
    \put(268,1998){\makebox(0,0)[l]{$C_{n(n+1)}$}}%
    \put(2151,581){\makebox(0,0)[l]{$T / J$}}%
    \put(1021,331){\makebox(0,0)[l]{$\Delta$}}%
    \put(631,518){\makebox(0,0){$8$}}%
    \put(843,494){\makebox(0,0){$6$}}%
    \put(1056,470){\makebox(0,0){$4$}}%
    \put(1268,445){\makebox(0,0){$2$}}%
    \put(1480,421){\makebox(0,0){$0$}}%
    \put(1693,397){\makebox(0,0){$-2$}}%
    \put(268,1998){\makebox(0,0)[l]{$C_{n(n+1)}$}}%
    \put(2151,581){\makebox(0,0)[l]{$T / J$}}%
    \put(1021,331){\makebox(0,0)[l]{$\Delta$}}%
    \put(322,1805){\makebox(0,0)[r]{$0.6$\hspace*{-0.25cm}}}%
    \put(322,1637){\makebox(0,0)[r]{$0.45$\hspace*{-0.25cm}}}%
    \put(322,1470){\makebox(0,0)[r]{$0.3$\hspace*{-0.25cm}}}%
    \put(322,1302){\makebox(0,0)[r]{$0.15$\hspace*{-0.25cm}}}%
    \put(322,1134){\makebox(0,0)[r]{$0$\hspace*{-0.25cm}}}%
    \put(1835,468){\makebox(0,0)[l]{$3$}}%
    \put(1989,602){\makebox(0,0)[l]{$2$}}%
    \put(2144,736){\makebox(0,0)[l]{$1$}}%
    \put(2299,870){\makebox(0,0)[l]{$0$}}%
    \put(631,518){\makebox(0,0){$8$}}%
    \put(843,494){\makebox(0,0){$6$}}%
    \put(1056,470){\makebox(0,0){$4$}}%
    \put(1268,445){\makebox(0,0){$2$}}%
    \put(1480,421){\makebox(0,0){$0$}}%
    \put(1693,397){\makebox(0,0){$-2$}}%
    \put(268,1998){\makebox(0,0)[l]{$C_{n(n+1)}$}}%
    \put(2151,581){\makebox(0,0)[l]{$T / J$}}%
    \put(1021,331){\makebox(0,0)[l]{$\Delta$}}%
    \end{picture}%
    \endgroup
    \vspace*{-1.0cm}
    \caption[$C_{n(n+1)}(\Delta, T / J)$ in the XXZ-model ($N = 4$, $J > 0$)]
            {\label{f_4C12} 
             The 3D-plot shows the concurrence \eq{C_{n(n+1)}} of the state of 
             two nearest neighbor qubits in the XXZ-model
             (\eq{N = 4}, \eq{J > 0}) 
             as a function of anisotropy \eq{\Delta} 
             and temperature \eq{T}.
             The 2D-plot shows the projection of critical temperature 
             \eq{T_{c}} (\ls{3}{0}) and lines of equal 
             \eq{C_{n(n+1)}} (\ls{1}{0}).
            } 
  \end{figure}

  The concurrence \eq{C_{n(n+1)}} of the state of two nearest neighbor 
  qubits as a function of anisotropy \eq{\Delta} and temperature \eq{T} 
  is depicted in Fig.~\ref{f_4C12}.%
  The energies together with the concurrences of the {\em individual} 
  eigenstates are responsible for all described features. 
  At $T=0$, the change of
  the ground state from \eq{E= \Delta} (\eq{s = \pm 2}, \eq{k = 0}) 
  to \eq{E= \mu_1} (\eq{s = 0}, \eq{k = 0}) causes the 
  discontinuity at \eq{\Delta = -1}. The position of the maximum in 
  \eq{C_{n(n+1)}(\Delta, T = 0)} is at \eq{\Delta = \Delta_{max} = 1}. 
  With increasing temperature, \eq{\Delta_{max}} increases but 
  \eq{C_{n(n+1)}(\Delta_{max}, T)} decreases monotonously.
  Concurrence \eq{C_{n(n+1)}} for fixed \eq{\Delta} is a 
  monotonously decreasing function of temperature. 
  As more energies 
  near the ground state energy exist as quicker decreases concurrence 
  with temperature. For example, the plateau region in the dependence of 
  \eq{C_{n(n+1)}} on \eq{T} for \eq{\Delta \gtrsim 4} stems from the
  with \eq{\Delta} increasing gap between ground state energy and most 
  energies of excited states. 
  The critical temperature 
  \eq{T_{c}} is defined as the lowest temperature above which the entanglement
  measure (here the concurrence) indicates an unentangled (part of the) state 
  (cf.~\cite[p.~155]{Nielsen:1998}). It is easily 
  identified as the intersection of the zero-surface and the surface of the 
  function \eq{C_{n(n+1)}} in Fig.~\ref{f_4C12}.
  The projection of the critical temperature \eq{T_c} and the lines of 
  equal \eq{C_{n(n+1)}} are depicted in the lower part 
  of Fig.~\ref{f_4C12}. In this way it is easy to identify parameter regions 
  of states with a certain minimal entanglement.
  Note that lines of finite equal
  concurrence are not increasing monotonously with increasing \eq{\Delta}
  but \eq{T_c} does. 

\begin{figure}[t]
    \sffamily
    \hspace*{0.25cm}%
    \centerline{\includegraphics[height=7.5cm, angle=-90]{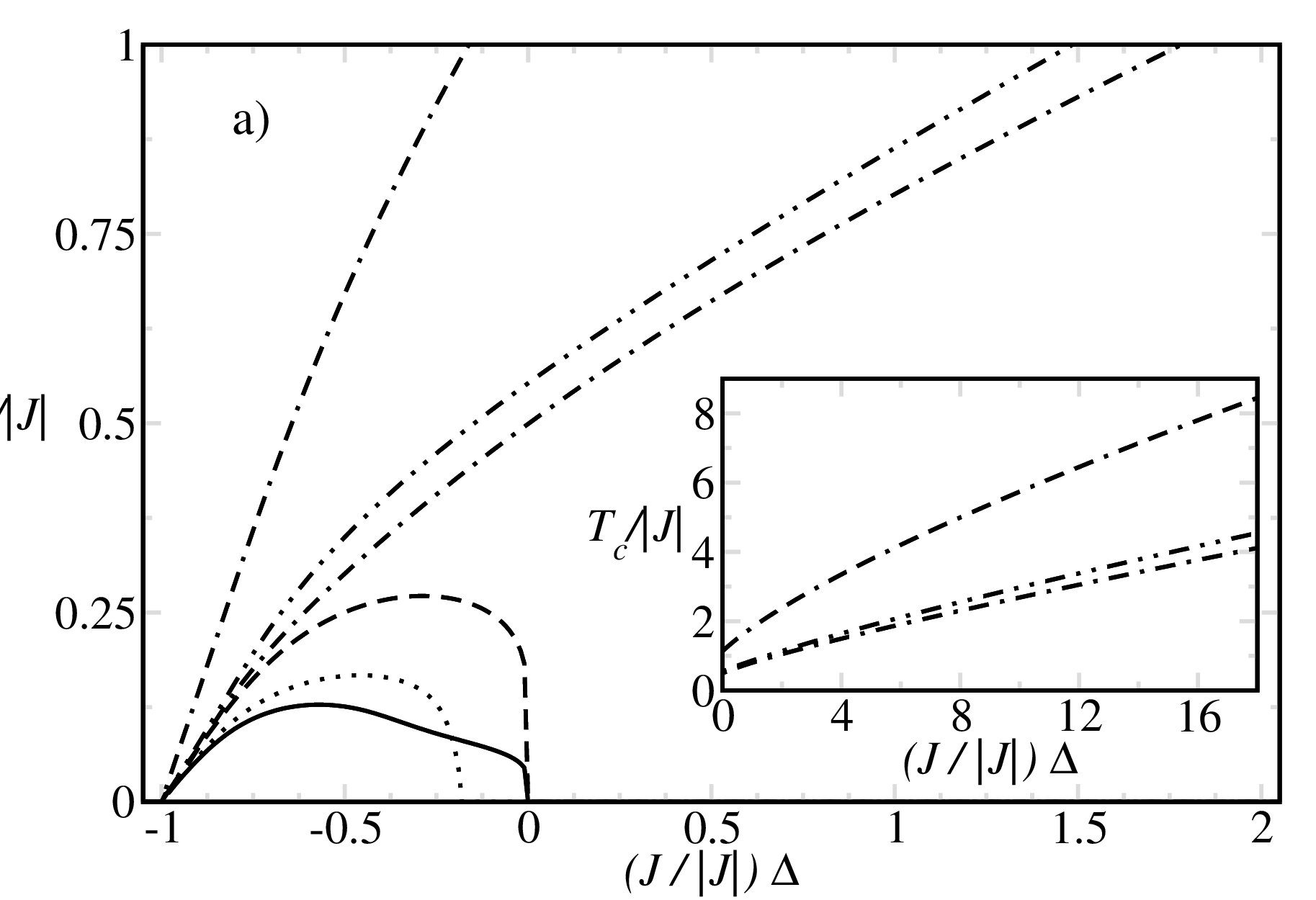}}
    \hspace*{0.25cm}%
    \centerline{\includegraphics[height=7.5cm, angle=-90]{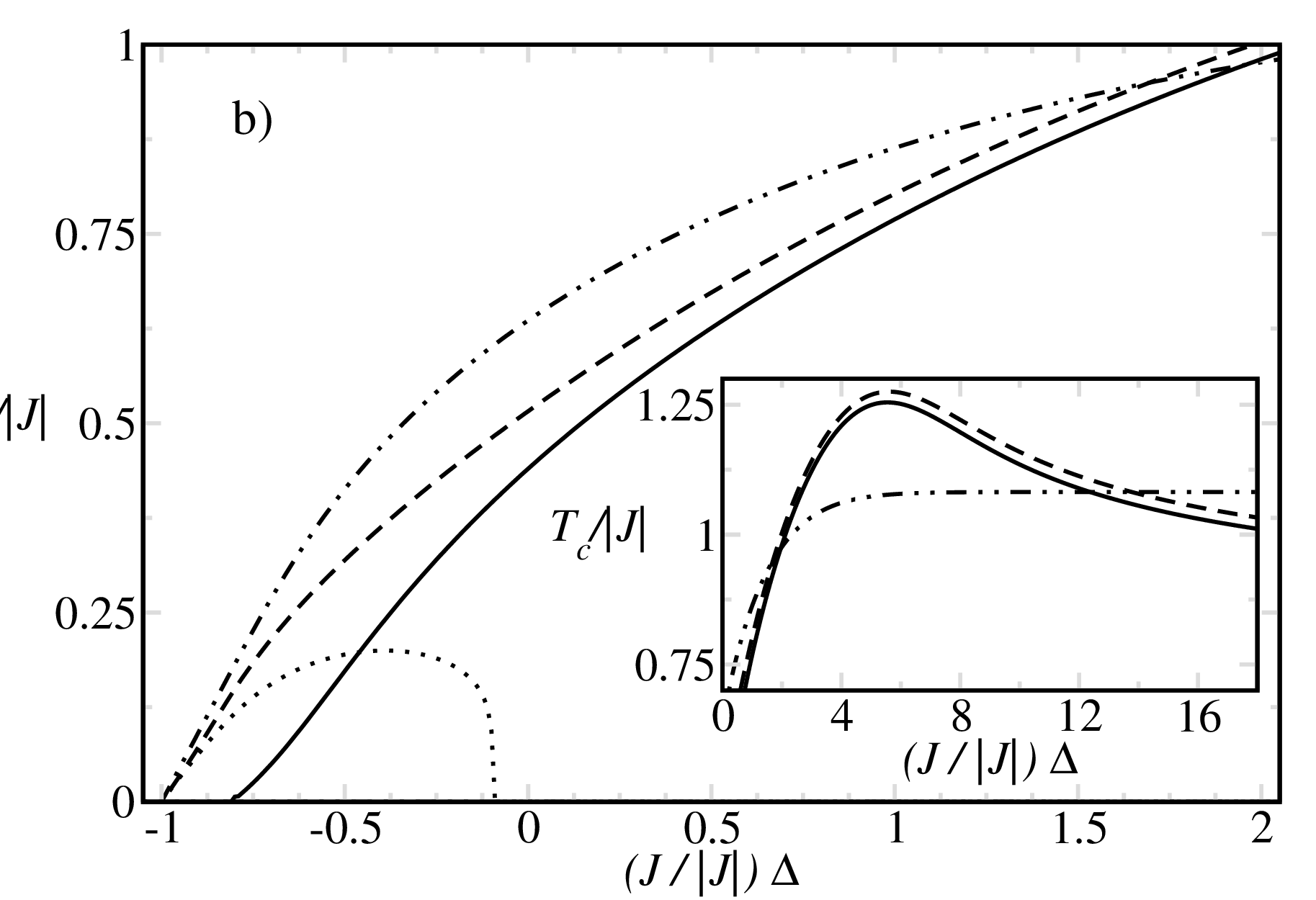}}
    \caption[$T_c(N, \Delta)$ of the XXZ-model ($J \lessgtr 0$)]
            {\label{f_Ring_Tc_2} 
             Critical temperature \eq{T_c} of the concurrence 
             of the state of two qubits in the XXZ-model (\eq{J \lessgtr 0}) 
             for \eq{2 \le N \le6} as a function of anisotropy \eq{\Delta}.
             Panel a) shows nearest neighbor qubits for \eq{N=2} (\ls{8}{0}), 
             \eq{N=4} (\ls{7}{0}) and \eq{N=6} (\ls{5}{0});
             next-to-nearest neighbor qubits for \eq{N=4} (\ls{3}{0}) 
             and \eq{N=6} (\ls{2}{0});
             next-to-next-to-nearest neighbor qubits for
             \eq{N=6} (\ls{1}{0}). 
             Panel b) displays  nearest neighbor qubits for
             \eq{N=3} (\eq{J < 0}: \ls{7}{0}; \eq{J > 0}: identical zero) 
             and 
             \eq{N=5} (\eq{J < 0}: \ls{3}{0}; \eq{J > 0}: \ls{1}{0}); 
             next-to-nearest neighbor qubits for 
             \eq{N=5} (\eq{J < 0}: \ls{2}{0}; \eq{J > 0}: identical zero).
             The insets give the dependence of these functions 
             at larger  values of \eq{(J / \abs{J}) \Delta}. Of course, 
             the entanglement vanishes in the Ising model limit of 
             (\ref{g_H}), i.e., for \eq{\abs{\Delta} \rightarrow \infty}.
            } 
  \end{figure}

  In Fig.~\ref{f_Ring_Tc_2}, the critical temperature \eq{T_c} of the 
  entanglement (measured in terms of concurrence) of the 
  state of two qubits in the XXZ-model (\eq{J \lessgtr 0}) 
  for \eq{2 \le N \le 6} as a function of anisotropy \eq{\Delta} is shown.%

  The transformation 
  \eq{J \rightarrow -J} and \eq{\Delta \rightarrow -\Delta} 
  leaves the critical temperature invariant for even \eq{N}.
  If \eq{\ket{\psi}} is an eigenstate of \eq{H(J, \Delta)} with
  eigenvalue \eq{E} then \eq{\ket{\phi} = A \ket{\psi}} is the 
  corresponding eigenstate of \eq{H(-J, -\Delta)} with the same
  eigenvalue and identical entanglement because \eq{A} is a local 
  unitary transformation 
  and entanglement is invariant under local unitary transformations. 
  Thus the thermal density operators of both Hamiltonians 
  are unitary equivalent and possess identical entanglement and 
  critical temperatures. 
  No such symmetry exists for odd \eq{N}. Actually, for the states of 
  nearest neighbor qubits (\eq{N = 3}) and next-to-nearest neighbor qubits 
  (\eq{N = 5}) entanglement is only possible for \eq{J < 0}. In all 
  considered cases the inequality 
  \eq{T_c(N, J < 0) \ge T_c(N, J > 0)} is valid.

  One observes in Fig.~\ref{f_Ring_Tc_2} 
  that \eq{T_{c} = 0} for \eq{(J / \abs{J}) \Delta \le -1} 
  independently of \eq{N} and the choice of the two qubits.
  It is known from \cite{Takahashi:1999} that 
  for all \eq{N}, \eq{J < 0} and \eq{\Delta \ge 1} 
  the two eigenstates of the Hamiltonian (\ref{g_H}) with 
  \eq{s = \pm \frac{N}{2}} are ground states. These ground states are not
  entangled and they cause the thermal state to be unentangled 
  for all temperatures. The same reasoning applies for even \eq{N}, 
  \eq{J > 0} and \eq{\Delta \le -1}, because of 
  the symmetries of the XXZ-model with periodic boundary conditions. 
  The ground state may change at different 
  \eq{\Delta} for odd \eq{N} and \eq{J > 0} 
  (e.g.~at \eq{\Delta \approx -0.809015} considering the XXZ-model
  with \eq{N = 5} and \eq{J > 0}).

  Furthermore critical temperature of geometrically equivalent aligned qubits
  is decreasing with increasing \eq{N} for even \eq{N}. This tendency is 
  consistent with the dependence of concurrence on \eq{N} in the isotropic
  Heisenberg model with an applied external magnetic field 
  (cf.~\cite{Arnesen:2001}).

\section{Analysis of an Experiment}
\label{k_Experiment}
  Finally, 
  the entanglement of the state of the quantum system in a NMR-experiment 
  about quantum error correction \cite{Knill:2001} is quantified in terms 
  of concurrence, i-concurrence and 3-tangle. 
  Five qubits are provided by different atoms in \eq{^{13}}C labeled 
  transcrotonic acid (synthesis and properties, see \cite{Knill:2000}) 
  solved in deuterated acetone. 

  One molecule can be approximately described by the one-dimensional 
  spatial inhomogeneous XXZ-model including an external magnetic field
  because the coupling constants of 
  non-neighboring qubits are much smaller than the coupling constants of 
  nearest neighbor qubits (see \cite{Knill:2001, Knill:2000}).
  The Hamiltonian \eq{H(J_n, \Delta, \omega_n)} of this model reads
  \eqxn{\label{g_H_exp}
        \begin{array}{rcl}
          H
          &=& \D \frac{1}{2} \sum\limits_{n=1}^{4} J_n 
                 \left( \sigma_{n}^{+} \sigma_{n+1}^{-} + 
                        \sigma_{n}^{-} \sigma_{n+1}^{+} + 
                        \frac{1}{2} \Delta \sigma_{n}^{z} \sigma_{n+1}^{z} 
                 \right) - \n
          & & \D \frac{1}{2} \sum\limits_{n=1}^{5} 
                 \omega_n \sigma_{n}^{z},
        \end{array}}
  where the coupling constants \eq{J_n} (\eq{n = 1, \ldots, 4}) specify the
  inhomogeneous strength of nearest neighbor interaction, \eq{\Delta} 
  determines the anisotropy in spin space and the effect of the external 
  magnetic field is included in \eq{\omega_n = \omega_n^p + \omega_n^c} 
  (\eq{n = 1, \ldots, 5}) which are the sums of precession frequencies 
  \eq{\omega_n^p} and chemical shifts \eq{\omega_n^c} for 
  each individual qubit (data in \cite{Knill:2001, Knill:2000}). 
  Of course, now open boundary conditions are applied.
 
\begin{figure}[b]
    \sffamily
    \centerline{\includegraphics[height=3.5cm, angle=0]{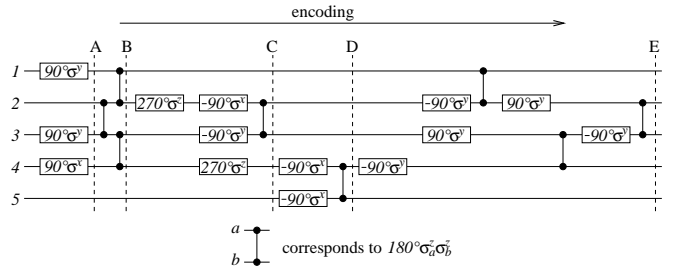}}
    \caption[Encoding based on the five-qubit code]
            {\label{f_code} 
             Encoding of qubit \eq{2} based on the five-qubit code. The
             horizontal lines represent the qubits. The gates denoted 
             \eq{\theta \sigma_{a}^{\alpha}} and 
             \eq{180^{\circ} \sigma_{a}^{z} \sigma_{b}^{z}} implement
             \eq{e^{-\frac{i}{2} \theta \sigma_{a}^{\alpha}}} and
             \eq{e^{-\frac{i}{4} \pi \sigma_{a}^{z} \sigma_{b}^{z}}}, 
             respectively. Here \eq{\alpha \in \{ x, y, z\} } and
             \eq{a, b \in \{ 1, 2, 3, 4, 5 \} }.
            } 
  \end{figure}

  The five-qubit code for quantum error correction is used to encode 
  qubit \eq{2} in the experiment. The encoding is shown in Fig.~\ref{f_code}.%
  The quantum system is in a highly mixed state, i.e., the coefficients of the 
  density operator are close to the coefficients of the identity operator, 
  because the experiment is performed at room temperature. 
  In the beginning, the quantum system is prepared in a way that only molecules 
  in the initial state \eq{\ket{11111}} give a signal on NMR measurements. 
  Then one says that the quantum system is in the {\em pseudo-pure} state 
  \eq{\ket{11111}} (Ref.~\cite{Gershenfeld:1997}). 
  The pseudo-pure state \eq{\ket{11111}} is an eigenstate of the 
  Hamiltonian (\ref{g_H_exp}) as well as the Hamiltonian including 
  {\em all} interactions of qubits and the applied external magnetic field 
  described in \cite{Knill:2001, Knill:2000}. Furthermore it is an eigenstate 
  of \eq{S^z}. Thus going to a frame of reference that rotates around the 
  \eq{z}-axis does not change the density operator of the initial state 
  (see \cite[p. 287]{Becker:2000}).

  The pseudo-pure state of the quantum system at several stages (A, B, C, D and
  E, cf.~Fig.~\ref{f_code}) during encoding was calculated by the 
  product-operator-formalism (see \cite[chapter 11]{Becker:2000}). 
  Therefore, the conservation of the pseudo-purity of the state of the 
  quantum system is assumed, i.e., there is no interaction
  between different molecules and encoding is implemented so quickly that no 
  decoherence occurs. The results are given in Table \ref{t_chain} together 
  with the expectation values \eq{K_{n}^{z}} and 
  \eq{K_{n}^{+} = \cc{ \left( K_{n}^{-} \right) }} 
  (with \eq{n = 1}, \eq{2}, \eq{\ldots}, \eq{5}).%
  \begin{table*}[t]
    \sffamily
    \caption[Pseudo-pure state at several states]
            {\label{t_chain}
             Pseudo-pure state \eq{\ket{\psi}} of the quantum system at 
             several states during encoding. The expectation values 
             \eq{K_{n}^{z}} and 
             \eq{K_{n}^{+} = \cc{ \left( K_{n}^{-} \right) }} 
             (with \eq{n = 1}, \eq{2}, \eq{\ldots}, \eq{5}) are 
             given for each state. Notation:
             \eq{\ket{1_x} := \frac{1}{\sqrt{2}} 
                              \left( \ket{1} + \ket{0} \right)},
             \eq{\ket{0_x} := \frac{1}{\sqrt{2}} 
                              \left( \ket{1} - \ket{0} \right)},
             \eq{\ket{1_y} := \frac{1}{\sqrt{2}} 
                              \left( \ket{1} + i \ket{0} \right)}, 
             \eq{\ket{0_y} := \frac{1}{\sqrt{2}} 
                              \left( \ket{1} - i \ket{0} \right)},
             \eq{\ket{1_z} := \ket{1}} and 
             \eq{\ket{0_z} := \ket{0}}.
            }
    \centering
    \newcommand{\pf}{\rule[-1.5ex]{0mm}{4.5ex}}
    \begin{tabular}[t]{|c|c|c|c|c|c|c|c|c|c|c|c|}
      \hline
      \pf Position
        & \eq{\ket{\psi}} 
        & \eq{K_1^z} 
        & \eq{K_2^z} 
        & \eq{K_3^z} 
        & \eq{K_4^z} 
        & \eq{K_5^z} 
        & \eq{K_1^+} 
        & \eq{K_2^+} 
        & \eq{K_3^+} 
        & \eq{K_4^+} 
        & \eq{K_5^+} \\
      \hline
      \hline
      \pf A
        & \eq{\ket{1_x 1_z 1_x 0_y 1_z}} 
        & \eq{0} 
        & \eq{1} 
        & \eq{0} 
        & \eq{0} 
        & \eq{1} 
        & \eq{\frac{1}{2}} 
        & \eq{0} 
        & \eq{\frac{1}{2}} 
        & \eq{-\frac{i}{2}} 
        & \eq{0} \\
      \hline
      \pf B
        & \eq{\frac{1}{\sqrt{2}} \ket{1_y 1_z} \otimes 
              \left( \ket{1_z 1_x} - \ket{0_z 0_x} \right) \otimes 
              \ket{1_z}} 
        & \eq{0} 
        & \eq{1} 
        & \eq{0} 
        & \eq{0} 
        & \eq{1} 
        & \eq{\frac{i}{2}} 
        & \eq{0} 
        & \eq{0} 
        & \eq{0} 
        & \eq{0} \\
      \hline
      \pf C
        & \eq{\frac{1}{\sqrt{2}} \ket{1_y} \otimes 
              \left( \ket{0_x 1_z 0_z} + 
                     \ket{1_x 0_z 1_z} \right) \otimes 
              \ket{1_z}} 
        & \eq{0} 
        & \eq{0} 
        & \eq{0} 
        & \eq{0} 
        & \eq{1} 
        & \eq{\frac{i}{2}} 
        & \eq{0} 
        & \eq{0} 
        & \eq{0} 
        & \eq{0} \\
      \hline
      \pf D
        & \eq{\frac{1}{2} \ket{1_y} \otimes 
              \left[ \ket{0_x 1_z} \otimes 
                     \left( \ket{1_z 0_x} + \ket{0_z 1_x} \right) - 
              i \ket{1_x 0_z} \otimes 
              \left( \ket{1_z 0_x} - \ket{0_z 1_x} \right) \right]} 
        & \eq{0} 
        & \eq{0} 
        & \eq{0} 
        & \eq{0} 
        & \eq{0} 
        & \eq{\frac{i}{2}} 
        & \eq{0} 
        & \eq{0} 
        & \eq{0} 
        & \eq{0} \\
      \hline
      \pf E
        & \begin{minipage}[c]{10cm}
            \pf
            \eq{\frac{1}{2 \sqrt{2}} \left\{ \ket{1_x} \otimes 
                \left[ \ket{0_y 1_z} \otimes 
                       \left( \ket{1_z 1_z} + \ket{0_z 0_z} \right) - 
                \ket{1_y 0_z} \otimes 
                \left( \ket{1_z 1_z} - \ket{0_z 0_z} \right) \right] 
                \right. +} \\ 
            \pf
            \eq{\quad \left. i \ket{0_x} \otimes 
                \left[ \ket{0_y 0_z} \otimes 
                       \left( \ket{1_z 0_z} + \ket{0_z 1_z} \right) - 
                \ket{1_y 1_z} \otimes 
                \left( \ket{1_z 0_z} - \ket{0_z 1_z} \right) \right] \right\}} 
          \end{minipage}
        & \eq{0} 
        & \eq{0} 
        & \eq{0} 
        & \eq{0} 
        & \eq{0} 
        & \eq{0} 
        & \eq{0} 
        & \eq{0} 
        & \eq{0} 
        & \eq{0} \\
      \hline
    \end{tabular}
  \end{table*}

  It is straightforward to calculate the entanglement of one qubit with 
  the remaining qubits by inserting these expectation values into
  Eq.~(\ref{g_IC_1r}). In this way it is easy to get a quick overview 
  about the possible entanglement in the quantum system. Note that it is not 
  appropriate to use Eqs.~(\ref{g_C}) and (\ref{g_CK}) or 
  (\ref{g_IC_2r}) here because the pseudo-pure state does 
  not comply with the necessary \eq{S^{z}}-symmetry in general.

  The pseudo-pure state at the various stages is now discussed in detail: 
  The initial state is not entangled. At position A, the state is not entangled 
  as well. So far only local operations have been performed and these cannot 
  create entanglement. 

  At position B, qubits \eq{1}, \eq{2} and \eq{5} are not
  entangled but \eq{C_{34} = 1}. Actually, the state of qubits \eq{3} 
  and \eq{4} at position B reads 
  \eq{\ket{\psi}_{34} = \frac{1}{2} \left( \ket{11} + \ket{10} - 
                                           \ket{01} + \ket{00} \right) }
  and it is conform to the Bell-states \eq{\ket{\psi^{\pm}}} and 
  \eq{\ket{\phi^\pm} = \frac{1}{\sqrt{2}} 
                       \left( \ket{00} \pm \ket{11} \right) } 
  up to a local unitary transformation.

  At position C, only qubits \eq{2}, \eq{3} and \eq{4} are entangled: 
  \eq{\bar{C}_{2-34} = \bar{C}_{3-24} = \bar{C}_{4-23} = \tau_{234} = 1}, 
  where the state of these qubits reads
  \eq{\ket{\psi}_{234} = \frac{1}{2} \left( \ket{110} + \ket{101} - 
                                           \ket{010} + \ket{001} \right) }.
  It is conform to the cat-state 
  \eq{\frac{1}{\sqrt{2}} \left( \ket{000} + \ket{111} \right) } 
  up to a local unitary transformation. Two out of these three qubits are
  not entangled as usual for a cat-state.

  At position D, only qubit \eq{1} is not entangled. The state of the 
  remaining qubits is conform to 
  \eq{\frac{1}{2} \left( \ket{0110} + \ket{0101} -
                         i \ket{1010} + i \ket{1001} \right)} 
  up to a local unitary transformation.
  The analysis of qubits 
  \eq{2}, \eq{3}, \eq{4} and \eq{5} shows no entanglement of the state of 
  two of these qubits. The entanglement of a state of three qubits 
  cannot be calculated because tracing off a qubit generates in general a 
  mixed state and i-concurrence can only be applied to pure states. 
  But it is \eq{\bar{C}_{2-345} = \bar{C}_{3-245} = 
  \bar{C}_{4-235} = \bar{C}_{5-234} = 1}, \eq{\bar{C}_{23-45} = 1} 
  and \eq{\bar{C}_{24-35} = \bar{C}_{25-34} = \sqrt{\frac{3}{2}}}.

  At the end of the encoding sequence (position E), all qubits are entangled: 
  \eq{\bar{C}_{A-B} = 1} if \eq{A} indicates one arbitrary qubit and \eq{B} the 
  remaining four qubits; \eq{\bar{C}_{A-B} = \sqrt{\frac{3}{2}}} if \eq{A} 
  indicates two arbitrary qubits and \eq{B} the remaining three qubits. 
  Again there is no entanglement of the state of two qubits and the 
  entanglement of a state of three or four qubits cannot be quantified so far.
  These results coincide with the ones in \cite{Meyer:2002}.
  It was already pointed out there that all states in a certain 
  fife-qubit error correction code subspace possess maximal global 
  entanglement but vanishing concurrences.

  Clearly, in this experiment, entanglement is created during encoding and it 
  expands in a geometrical sense, i.e., the number of qubits involved in 
  the entanglement increases with the progressing encoding sequence.

  Unfortunately, it is not possible to quantify the entanglement of the state 
  at positions D and E completely because of the lack of suitable measures. 
  But all calculated i-concurrences exhibit their maximal values at position E.
  Thus it is a reasonable conjecture that an entanglement of four or less 
  qubits does not exist there because entanglement cannot be shared arbitrarily 
  (cf. \cite{Coffman:2000}).

\section{Summary}
\label{k_Summary}

  The entanglement measures concurrence, i-concurrence (for one
  or two qubits in one subsystem) and 3-tangle have been successfully 
  expressed in terms of correlation functions. In addition, 
  necessary and sufficient conditions for a positive concurrence 
  have been formulated. 
  These results have been used in the remaining paper because they
  can simplify calculations: The concurrence of 
  eigenstates or the thermal state have been calculated analytically knowing 
  only the energies of the eigenstates and their dependences on the
  parameters of the system. Furthermore potential quantum entanglement in a 
  quantum system has been detected by the examination of spin expectation
  values.
  
  A detailed analysis of concurrence and critical temperature in the 
  XXZ-model with \eq{2 \le N \le 6} qubits has been accomplished.

  Finally, the entanglement of the state in a NMR-experiment has been
  discussed quantitatively. Different kinds of entanglement 
  have been identified. This calculation shows the relevance of 
  entanglement measures in actual experiments because they allow an
  analysis of the importance of entanglement for the quantum-algorithms.
  Despite the information, which is obtained with the available measures,
  further measures are needed for a complete insight. 

  The entanglement measures might be useful designing
  new experiments (possibly utilizing advanced types
  of qubits, e.g., spin cluster qubits \cite{Meier:2003}) that set up 
  states with different entanglement and prove or disprove the benefit 
  of entanglement in different quantum-algorithms.

  One of us (H.F.) thanks John Schliemann for useful discussions.



\begin{thebibliography}{32}
\expandafter\ifx\csname natexlab\endcsname\relax\def\natexlab#1{#1}\fi
\expandafter\ifx\csname bibnamefont\endcsname\relax
  \def\bibnamefont#1{#1}\fi
\expandafter\ifx\csname bibfnamefont\endcsname\relax
  \def\bibfnamefont#1{#1}\fi
\expandafter\ifx\csname url\endcsname\relax
  \def\url#1{\texttt{#1}}\fi
\expandafter\ifx\csname urlprefix\endcsname\relax\def\urlprefix{URL }\fi
\providecommand*{\bibinfo}[2]{#2}
\providecommand*{\eprint}[1]{#1}
\providecommand*{\url}[1]{#1}
\begingroup\makeatletter
 \@temptokena{%
  \expandafter\ifx\csname citenamefont\endcsname\relax
   \DeclareRobustCommand\citenamefont{\@firstofone}%
   \global\let\citenamefont\citenamefont
   \global\expandafter\let\csname citenamefont \expandafter\endcsname\csname
  citenamefont \endcsname
  \fi
 }\if@filesw\immediate\write\@auxout{\the\@temptokena}\fi
\expandafter\endgroup\the\@temptokena

\bibitem[{\citenamefont{Schr\"odinger}(1935)}]{Schroedinger:1935x}
\bibinfo{author}{\bibfnamefont{E.}~\bibnamefont{Schr\"odinger}},
  \bibinfo{journal}{Die Naturwissenschaften} \textbf{\bibinfo{volume}{23}},
  \bibinfo{pages}{807, 823, 844} (\bibinfo{year}{1935}).

\bibitem[{\citenamefont{Macchiavello}
  \emph{et~al.}(2000)\citenamefont{Macchiavello, Palma, and
  Zeilinger}}]{Macchiavello:2000}
\bibinfo{editor}{\bibfnamefont{C.}~\bibnamefont{Macchiavello}},
  \bibinfo{editor}{\bibfnamefont{G.~M.} \bibnamefont{Palma}}, \bibnamefont{and}
  \bibinfo{editor}{\bibfnamefont{A.}~\bibnamefont{Zeilinger}}, eds.,
  \emph{\bibinfo{title}{Quantum Computation and Quantum Information Theory}}
  (\bibinfo{publisher}{World Scientific Publishing Co.~Pte.~Ltd.},
  \bibinfo{year}{2000}).

\bibitem[{\citenamefont{Nielsen and Chuang}(2001)}]{Nielsen:2001}
\bibinfo{author}{\bibfnamefont{M.~A.} \bibnamefont{Nielsen}} \bibnamefont{and}
  \bibinfo{author}{\bibfnamefont{I.~L.} \bibnamefont{Chuang}},
  \emph{\bibinfo{title}{Quantum Computation and Quantum Information}}
  (\bibinfo{publisher}{Cambridge University Press}, \bibinfo{year}{2001}).

\bibitem[{\citenamefont{O'Connor and Wootters}(2001)}]{Connor:2001}
\bibinfo{author}{\bibfnamefont{K.~M.} \bibnamefont{O'Connor}} \bibnamefont{and}
  \bibinfo{author}{\bibfnamefont{W.~K.} \bibnamefont{Wootters}},
  \bibinfo{journal}{Phys. Rev. A}
  \textbf{\bibinfo{volume}{63}}(\bibinfo{number}{5}), \bibinfo{pages}{052302}
  (\bibinfo{year}{2001}).

\bibitem[{\citenamefont{Arnesen} \emph{et~al.}(2001)\citenamefont{Arnesen,
  Bose, and Vedral}}]{Arnesen:2001}
\bibinfo{author}{\bibfnamefont{M.~C.} \bibnamefont{Arnesen}},
  \bibinfo{author}{\bibfnamefont{S.}~\bibnamefont{Bose}}, \bibnamefont{and}
  \bibinfo{author}{\bibfnamefont{V.}~\bibnamefont{Vedral}},
  \bibinfo{journal}{Phys. Rev. Lett.}
  \textbf{\bibinfo{volume}{87}}(\bibinfo{number}{1}), \bibinfo{pages}{017901}
  (\bibinfo{year}{2001}).

\bibitem[{\citenamefont{Wang} \emph{et~al.}(2001)\citenamefont{Wang, Fu, and
  Solomon}}]{Wang:2001_12}
\bibinfo{author}{\bibfnamefont{X.}~\bibnamefont{Wang}},
  \bibinfo{author}{\bibfnamefont{H.}~\bibnamefont{Fu}}, \bibnamefont{and}
  \bibinfo{author}{\bibfnamefont{A.~I.} \bibnamefont{Solomon}},
  \bibinfo{journal}{J. Phys. A}
  \textbf{\bibinfo{volume}{34}}(\bibinfo{number}{50}), \bibinfo{pages}{11307}
  (\bibinfo{year}{2001}).

\bibitem[{\citenamefont{Wang}(2002{\natexlab{a}})}]{Wang:2002_3}
\bibinfo{author}{\bibfnamefont{X.}~\bibnamefont{Wang}}, \bibinfo{journal}{New
  J. Phys.} \textbf{\bibinfo{volume}{4}}, \bibinfo{pages}{11}
  (\bibinfo{year}{2002}{\natexlab{a}}).

\bibitem[{\citenamefont{Wang and Zanardi}(2002)}]{Wang:2002_8}
\bibinfo{author}{\bibfnamefont{X.}~\bibnamefont{Wang}} \bibnamefont{and}
  \bibinfo{author}{\bibfnamefont{P.}~\bibnamefont{Zanardi}},
  \bibinfo{journal}{Phys. Lett. A}
  \textbf{\bibinfo{volume}{301}}(\bibinfo{number}{1-2}), \bibinfo{pages}{1}
  (\bibinfo{year}{2002}).

\bibitem[{\citenamefont{Wang}(2002{\natexlab{b}})}]{Wang:2002_10}
\bibinfo{author}{\bibfnamefont{X.}~\bibnamefont{Wang}}, \bibinfo{journal}{Phys.
  Rev. A} \textbf{\bibinfo{volume}{66}}(\bibinfo{number}{4}),
  \bibinfo{pages}{044305} (\bibinfo{year}{2002}{\natexlab{b}}).

\bibitem[{\citenamefont{Schliemann}(2002)}]{Schliemann:2002}
\bibinfo{author}{\bibfnamefont{J.}~\bibnamefont{Schliemann}},
  \bibinfo{journal}{quant-ph/0212114}  (\bibinfo{year}{2002}).

\bibitem[{\citenamefont{Gunlycke} \emph{et~al.}(2001)\citenamefont{Gunlycke,
  Kendon, Vedral, and Bose}}]{Gunlycke:2001}
\bibinfo{author}{\bibfnamefont{D.}~\bibnamefont{Gunlycke}},
  \bibinfo{author}{\bibfnamefont{V.~M.} \bibnamefont{Kendon}},
  \bibinfo{author}{\bibfnamefont{V.}~\bibnamefont{Vedral}}, \bibnamefont{and}
  \bibinfo{author}{\bibfnamefont{S.}~\bibnamefont{Bose}},
  \bibinfo{journal}{Phys. Rev. A}
  \textbf{\bibinfo{volume}{64}}(\bibinfo{number}{4}), \bibinfo{pages}{042302}
  (\bibinfo{year}{2001}).

\bibitem[{\citenamefont{Kamta and Starace}(2002)}]{Kamta:2002}
\bibinfo{author}{\bibfnamefont{G.~L.} \bibnamefont{Kamta}} \bibnamefont{and}
  \bibinfo{author}{\bibfnamefont{A.~F.} \bibnamefont{Starace}},
  \bibinfo{journal}{Phys. Rev. Lett.}
  \textbf{\bibinfo{volume}{88}}(\bibinfo{number}{10}), \bibinfo{pages}{107901}
  (\bibinfo{year}{2002}).

\bibitem[{\citenamefont{Osterloh} \emph{et~al.}(2002)\citenamefont{Osterloh,
  Amico, Falci, and Fazio}}]{Osterloh:2002}
\bibinfo{author}{\bibfnamefont{A.}~\bibnamefont{Osterloh}},
  \bibinfo{author}{\bibfnamefont{L.}~\bibnamefont{Amico}},
  \bibinfo{author}{\bibfnamefont{G.}~\bibnamefont{Falci}}, \bibnamefont{and}
  \bibinfo{author}{\bibfnamefont{R.}~\bibnamefont{Fazio}},
  \bibinfo{journal}{Nature} \textbf{\bibinfo{volume}{416}},
  \bibinfo{pages}{608} (\bibinfo{year}{2002}).

\bibitem[{\citenamefont{Osborne and Nielsen}(2002)}]{Osborne:2002}
\bibinfo{author}{\bibfnamefont{T.~J.} \bibnamefont{Osborne}} \bibnamefont{and}
  \bibinfo{author}{\bibfnamefont{M.~A.} \bibnamefont{Nielsen}},
  \bibinfo{journal}{Phys. Rev. A}
  \textbf{\bibinfo{volume}{66}}(\bibinfo{number}{3}), \bibinfo{pages}{032110}
  (\bibinfo{year}{2002}).

\bibitem[{\citenamefont{Bose and Chattopadhyay}(2002)}]{Bose:2002}
\bibinfo{author}{\bibfnamefont{I.}~\bibnamefont{Bose}} \bibnamefont{and}
  \bibinfo{author}{\bibfnamefont{E.}~\bibnamefont{Chattopadhyay}},
  \bibinfo{journal}{Phys. Rev. A}
  \textbf{\bibinfo{volume}{66}}(\bibinfo{number}{6}), \bibinfo{pages}{062320}
  (\bibinfo{year}{2002}).

\bibitem[{\citenamefont{Vidal} \emph{et~al.}(2002)\citenamefont{Vidal, Latorre,
  Rico, and Kitaev}}]{Vidal:2002}
\bibinfo{author}{\bibfnamefont{G.}~\bibnamefont{Vidal}},
  \bibinfo{author}{\bibfnamefont{J.~I.} \bibnamefont{Latorre}},
  \bibinfo{author}{\bibfnamefont{E.}~\bibnamefont{Rico}}, \bibnamefont{and}
  \bibinfo{author}{\bibfnamefont{A.}~\bibnamefont{Kitaev}},
  \bibinfo{journal}{quant-ph/0211074}  (\bibinfo{year}{2002}).

\bibitem[{\citenamefont{Latorre} \emph{et~al.}(2003)\citenamefont{Latorre,
  Rico, and Vidal}}]{Latorre:2003}
\bibinfo{author}{\bibfnamefont{J.~I.} \bibnamefont{Latorre}},
  \bibinfo{author}{\bibfnamefont{E.}~\bibnamefont{Rico}}, \bibnamefont{and}
  \bibinfo{author}{\bibfnamefont{G.}~\bibnamefont{Vidal}},
  \bibinfo{journal}{quant-ph/0304098}  (\bibinfo{year}{2003}).

\bibitem[{\citenamefont{Meyer and Wallach}(2002)}]{Meyer:2002}
\bibinfo{author}{\bibfnamefont{D.~A.} \bibnamefont{Meyer}} \bibnamefont{and}
  \bibinfo{author}{\bibfnamefont{N.~R.} \bibnamefont{Wallach}},
  \bibinfo{journal}{J. Math. Phys.}
  \textbf{\bibinfo{volume}{43}}(\bibinfo{number}{9}), \bibinfo{pages}{4273}
  (\bibinfo{year}{2002}).

\bibitem[{\citenamefont{Hill and Wootters}(1997)}]{Hill:1997}
\bibinfo{author}{\bibfnamefont{S.}~\bibnamefont{Hill}} \bibnamefont{and}
  \bibinfo{author}{\bibfnamefont{W.~K.} \bibnamefont{Wootters}},
  \bibinfo{journal}{Phys. Rev. Lett.}
  \textbf{\bibinfo{volume}{78}}(\bibinfo{number}{26}), \bibinfo{pages}{5022}
  (\bibinfo{year}{1997}).

\bibitem[{\citenamefont{Wootters}(1998)}]{Wootters:1998}
\bibinfo{author}{\bibfnamefont{W.~K.} \bibnamefont{Wootters}},
  \bibinfo{journal}{Phys. Rev. Lett.}
  \textbf{\bibinfo{volume}{80}}(\bibinfo{number}{10}), \bibinfo{pages}{2245}
  (\bibinfo{year}{1998}).

\bibitem[{\citenamefont{Rungta} \emph{et~al.}(2001)\citenamefont{Rungta,
  Bu\v{z}ek, Caves, Hillery, and Milburn}}]{Rungta:2001}
\bibinfo{author}{\bibfnamefont{P.}~\bibnamefont{Rungta}},
  \bibinfo{author}{\bibfnamefont{V.}~\bibnamefont{Bu\v{z}ek}},
  \bibinfo{author}{\bibfnamefont{C.~M.} \bibnamefont{Caves}},
  \bibinfo{author}{\bibfnamefont{M.}~\bibnamefont{Hillery}}, \bibnamefont{and}
  \bibinfo{author}{\bibfnamefont{G.~J.} \bibnamefont{Milburn}},
  \bibinfo{journal}{Phys. Rev. A}
  \textbf{\bibinfo{volume}{64}}(\bibinfo{number}{4}), \bibinfo{pages}{042315}
  (\bibinfo{year}{2001}).

\bibitem[{\citenamefont{Coffman} \emph{et~al.}(2000)\citenamefont{Coffman,
  Kundu, and Wootters}}]{Coffman:2000}
\bibinfo{author}{\bibfnamefont{V.}~\bibnamefont{Coffman}},
  \bibinfo{author}{\bibfnamefont{J.}~\bibnamefont{Kundu}}, \bibnamefont{and}
  \bibinfo{author}{\bibfnamefont{W.~K.} \bibnamefont{Wootters}},
  \bibinfo{journal}{Phys. Rev. A}
  \textbf{\bibinfo{volume}{61}}(\bibinfo{number}{5}), \bibinfo{pages}{052306}
  (\bibinfo{year}{2000}).

\bibitem[{\citenamefont{Knill} \emph{et~al.}(2001)\citenamefont{Knill,
  Laflamme, Martinez, and Negrevergne}}]{Knill:2001}
\bibinfo{author}{\bibfnamefont{E.}~\bibnamefont{Knill}},
  \bibinfo{author}{\bibfnamefont{R.}~\bibnamefont{Laflamme}},
  \bibinfo{author}{\bibfnamefont{R.}~\bibnamefont{Martinez}}, \bibnamefont{and}
  \bibinfo{author}{\bibfnamefont{C.}~\bibnamefont{Negrevergne}},
  \bibinfo{journal}{Phys. Rev. Lett.}
  \textbf{\bibinfo{volume}{86}}(\bibinfo{number}{25}), \bibinfo{pages}{5811}
  (\bibinfo{year}{2001}).

\bibitem[{\citenamefont{Takahashi}(1999)}]{Takahashi:1999}
\bibinfo{author}{\bibfnamefont{M.}~\bibnamefont{Takahashi}},
  \emph{\bibinfo{title}{Thermodynamics of one-dimensional solvable models}}
  (\bibinfo{publisher}{Cambridge University Press}, \bibinfo{year}{1999}).

\bibitem[{\citenamefont{Yang and Yang}(1966)}]{Yang:1966}
\bibinfo{author}{\bibfnamefont{C.~N.} \bibnamefont{Yang}} \bibnamefont{and}
  \bibinfo{author}{\bibfnamefont{C.~P.} \bibnamefont{Yang}},
  \bibinfo{journal}{Phys. Rev.}
  \textbf{\bibinfo{volume}{147}}(\bibinfo{number}{1}), \bibinfo{pages}{303}
  (\bibinfo{year}{1966}).

\bibitem[{\citenamefont{Orbach}(1958)}]{Orbach:1958}
\bibinfo{author}{\bibfnamefont{R.}~\bibnamefont{Orbach}},
  \bibinfo{journal}{Phys. Rev.}
  \textbf{\bibinfo{volume}{112}}(\bibinfo{number}{2}), \bibinfo{pages}{309}
  (\bibinfo{year}{1958}).

\bibitem[{\citenamefont{Stroganov}(2001)}]{Stroganov:2001}
\bibinfo{author}{\bibfnamefont{Y.}~\bibnamefont{Stroganov}},
  \bibinfo{journal}{J. Phys. A}
  \textbf{\bibinfo{volume}{34}}(\bibinfo{number}{13}), \bibinfo{pages}{L179}
  (\bibinfo{year}{2001}).

\bibitem[{\citenamefont{Nielsen}(1998)}]{Nielsen:1998}
\bibinfo{author}{\bibfnamefont{M.~A.} \bibnamefont{Nielsen}},
  \emph{\bibinfo{title}{Quantum Information Theory}}, Ph.D. thesis,
  \bibinfo{school}{The University of New Mexico Albuquerque}
  (\bibinfo{year}{1998}), \bibinfo{note}{quant-ph/0011036}.

\bibitem[{\citenamefont{Knill} \emph{et~al.}(2000)\citenamefont{Knill,
  Laflamme, Martinez, and Tseng}}]{Knill:2000}
\bibinfo{author}{\bibfnamefont{E.}~\bibnamefont{Knill}},
  \bibinfo{author}{\bibfnamefont{R.}~\bibnamefont{Laflamme}},
  \bibinfo{author}{\bibfnamefont{R.}~\bibnamefont{Martinez}}, \bibnamefont{and}
  \bibinfo{author}{\bibfnamefont{C.-H.} \bibnamefont{Tseng}},
  \bibinfo{journal}{Nature} \textbf{\bibinfo{volume}{404}},
  \bibinfo{pages}{368} (\bibinfo{year}{2000}).

\bibitem[{\citenamefont{Gershenfeld and Chuang}(1997)}]{Gershenfeld:1997}
\bibinfo{author}{\bibfnamefont{N.~A.} \bibnamefont{Gershenfeld}}
  \bibnamefont{and} \bibinfo{author}{\bibfnamefont{I.~L.}
  \bibnamefont{Chuang}}, \bibinfo{journal}{Science}
  \textbf{\bibinfo{volume}{275}}, \bibinfo{pages}{350} (\bibinfo{year}{1997}).

\bibitem[{\citenamefont{Becker}(2000)}]{Becker:2000}
\bibinfo{author}{\bibfnamefont{E.~D.} \bibnamefont{Becker}},
  \emph{\bibinfo{title}{High Resolution {NMR}}} (\bibinfo{publisher}{Academic
  Press}, \bibinfo{year}{2000}).

\bibitem[{\citenamefont{Meier} \emph{et~al.}(2003)\citenamefont{Meier, Levy,
  and Loss}}]{Meier:2003}
\bibinfo{author}{\bibfnamefont{F.}~\bibnamefont{Meier}},
  \bibinfo{author}{\bibfnamefont{J.}~\bibnamefont{Levy}}, \bibnamefont{and}
  \bibinfo{author}{\bibfnamefont{D.}~\bibnamefont{Loss}},
  \bibinfo{journal}{Phys. Rev. Lett.}
  \textbf{\bibinfo{volume}{90}}(\bibinfo{number}{4}), \bibinfo{pages}{047901}
  (\bibinfo{year}{2003}).

\end{thebibliography}
\end{document}